\documentclass[fleqn,twoside]{article}
\usepackage{amsmath,amssymb,espcrc2,color}

\usepackage{graphicx}
\usepackage[figuresright]{rotating}
\newcommand{\url}[1]{{\tt #1}}
\newcommand{\lsim}
{\;\raisebox{-.3em}{$\stackrel{\displaystyle <}{\sim}$}\;}

\newcommand{\gmt}{\ensuremath{(g-2)_\mu}}

\newcommand{\MW}{M_W}
\newcommand{\MZ}{M_Z}
\newcommand{\Mh}{M_h}

\newcommand{\MHSM}{M_H^{\rm SM}}
\newcommand{\mt}{m_t}

\newcommand{\gev}{\,\, \mathrm{GeV}}

\definecolor{Orange}{named}{Orange}
\definecolor{Purple}{named}{Purple}

\graphicspath{{figs/}}

\title{Predictions for $\mt$ and $\MW$
       in Minimal Supersymmetric Models} 

\author{
O.~Buchmueller\address[Imperial]
   {High\,Energy\,Physics\,Group, Imperial\,College, Blackett\,Lab., 
    Prince\,Consort\,Road, London\,SW7\,2AZ,\,UK},
R.~Cavanaugh\address[FNAL]
   {Fermi National Accelerator Laboratory, P.O. Box 500, 
    Batavia, Illinois 60510, USA}\hbox{$^{\rm ,}$}\address[UIC]
   {Physics Department, University of Illinois at Chicago, Chicago, 
    Illinois 60607-7059, USA},
A.~De Roeck\address[CERN]
   {CERN, CH--1211 Gen\`eve 23, Switzerland}\hbox{$^{\rm ,}$}\address[Antwerpen]
   {Antwerp University, B--2610 Wilrijk, Belgium},
J.R.~Ellis\addressmark[CERN],
H.~Fl\"acher\address[Rochester]
   {Department of Physics and Astronomy, University of Rochester, 
    Rochester, New York 14627, USA},
S.~Heinemeyer\address[Santander]
   {Instituto de F\'{\i}sica de Cantabria (CSIC-UC), 
    E--39005 Santander, Spain},
G.~Isidori\address[Frascati]
   {INFN, Laboratori Nazionali di Frascati, Via E. Fermi 40, 
    I--00044 Frascati, Italy}\hbox{$^{\rm ,}$}\address[IAS]
   {Institute for Advanced Study, T.~U.~M\"unchen, 
    Arcisstra\ss e 21, D-80333 M\"unchen, Germany},  
K.A.~Olive\address[Minnesota] 
   {William I. Fine Theoretical Physics Institute, Univ.\ of Minnesota,
     Minneapolis, Minnesota 55455,\,USA}, 
F.J.~Ronga\address[ETHZ]
   {Institute for Particle Physics, ETH Z\"urich, CH--8093 Z\"urich, 
   Switzerland},
G.~Weiglein\address[DESY]
   {DESY, Notkestrasse 85, D--22603 Hamburg, Germany}
}
       
\begin{document}

\begin{abstract}
Using a frequentist analysis of experimental constraints within two versions of the 
minimal supersymmetric extension of the Standard Model, 
we derive the predictions for the top quark mass, $\mt$, and the
$W$~boson mass, $\MW$. 
We find that the supersymmetric predictions for both $\mt$ and $\MW$, 
obtained by incorporating all the relevant experimental information
and state-of-the-art theoretical predictions,  
are highly compatible with the experimental values with 
small remaining uncertainties, yielding an improvement compared to the
case of the Standard Model.

% Preprint numbers
\bigskip
\begin{center}
CERN-PH-TH/2009-223, DESY 09-207, FTPI-MINN-09/44,
UMN-TH-2827/09
\end{center}
%\vspace{-0.5cm}
\end{abstract}

\maketitle

%---------------------------------------------------------------------
%\section{Introduction}
%\label{sec:intro}
%---------------------------------------------------------------------

One of the most impressive successes of the Standard Model (SM) has
been the accurate prediction of the mass of the top quark obtained
from a fit 
to precision electroweak measurements at LEP and the SLC~\cite{lepewwg}, 
which agrees very well with the value measured at
the Tevatron~\cite{mt1731}.  To this may be added the equally successful
prediction of the $W$ mass~\cite{lepewwg,MW80399pm023}. 
The successes of these comparisons between theory and experiment
require the incorporation of higher-order quantum corrections. In the SM
these receive contributions from the postulated
Higgs boson. Indeed, the precision data favour
a relatively light Higgs boson weighing
$\lsim 150 \gev$~\cite{lepewwg}.
 
One theoretical framework that predicts such a light Higgs boson is
supersymmetry (SUSY)~\cite{Haber:1984rc},
which also possesses the ability to render
more natural the electroweak mass hierarchy, contains a
plausible candidate for astrophysical  dark matter, facilitates
grand unification, and offers a possible explanation of the apparent discrepancy 
between the experimental measurement of the anomalous magnetic
moment of the muon, \gmt, and the theoretical value calculated
within the SM. There have been many analyses of the
possible masses of particles within the minimal supersymmetric
extension of the Standard Model (MSSM), taking into account the
experimental, phenomenological and astrophysical constraints. For
example, we have presented sparticle mass
predictions~\cite{Master1,Master2,Master3} on the basis 
of a frequentist analyses of the relevant constraints in the context of
simple models for SUSY breaking such as the CMSSM (in
which the input scalar masses $m_0$, gaugino masses $m_{1/2}$ and
soft trilinear parameters $A_0$ are each universal at the GUT scale)
and the NUHM1 (in which a common SUSY-breaking contribution
to the Higgs masses is allowed to be non-universal). 
For an extensive list of references, see~\cite{Master3}.

These analyses favour relatively light masses for the sparticles, 
indicating significant sensitivity of the precision observables to 
quantum effects of supersymmetric particles.
It is therefore desirable to revisit the successful predictions of the
SM, in particular the show-case predictions of $\mt$ and
$\MW$, to see how they are affected in the CMSSM and NUHM1.
In particular, one may ask whether the SM prediction
of $\mt$ and $\MW$ is improved, relaxed or otherwise altered in these models.
The answer to this key question is non-trivial, since low-mass sparticles such as the
${\tilde t}$ and ${\tilde b}$ may contribute significantly to the
prediction of electroweak observables~\cite{PomssmRep}, and the (lightest)
Higgs mass is no longer an independent quantity, but also depends on the
sparticle masses as we discuss below.

In this Letter, we make supersymmetric predictions for $\mt$ and 
$\MW$ within the same framework as our previous
frequentist analyses of the CMSSM and NUHM1 parameter spaces~\cite{Master1,Master2,Master3}.
The treatments of the experimental, phenomenological and astrophysical constraints
are nearly identical with those in~\cite{Master3}. Here,
we employ the updated SM value of \gmt\ which
includes a new set of
low-energy $e^+ e^-$ data~\cite{g-2babar}.
The new value of \gmt\ \cite{newg-2} does not significantly alter the
regions of the CMSSM and NUHM1 parameter spaces favoured in our
previous analyses.

Our statistical treatment of the CMSSM and NUHM1 makes use of  
a large sample of points (about $3 \times 10^6$) 
in the SUSY parameter spaces 
obtained with the Markov Chain Monte Carlo
(MCMC) technique~\cite{sedews}.
Our analysis is entirely frequentist, and avoids any
ambiguity associated with the choices of Bayesian priors.
The evaluations are performed using the 
{\tt MasterCode}~\cite{Master1,Master2,Master3,MasterWWW}, 
which includes the following theoretical codes. For the RGE running of
the soft SUSY-breaking parameters, it uses
{\tt SoftSUSY}~\cite{Allanach:2001kg}, which is combined consistently
with the codes used for the various low-energy observables:
{\tt FeynHiggs}~\cite{Heinemeyer:1998yj,mhiggsAEC,Frank:2006yh}  
is used for the evaluation of the Higgs masses and  
$a_\mu^{\rm SUSY}$  (see also
\cite{Moroi:1995yh,PomssmRep}),
for the other electroweak precision data we have included 
a code based on~\cite{Heinemeyer:2006px,Heinemeyer:2007bw},
{\tt SuFla}~\cite{Isidori:2006pk,Isidori:2007jw} and 
{\tt SuperIso}~\cite{Mahmoudi:2008tp,Eriksson:2008cx}
are used for flavour-related observables, 
and for dark-matter-related observables
{\tt MicrOMEGAs}~\cite{Belanger:2006is} and
{\tt DarkSUSY}~\cite{Gondolo:2005we} are used.
In the combination of the various codes,
{\tt MasterCode} makes extensive use of the SUSY
Les Houches Accord~\cite{Skands:2003cj,Allanach:2008qq}.
 
\begin{figure}[htb!]
%%%%%%%%%%%%%%%%%%%%%%%%%%%%%%%
\begin{center}
{\resizebox{7.5cm}{!}{\includegraphics{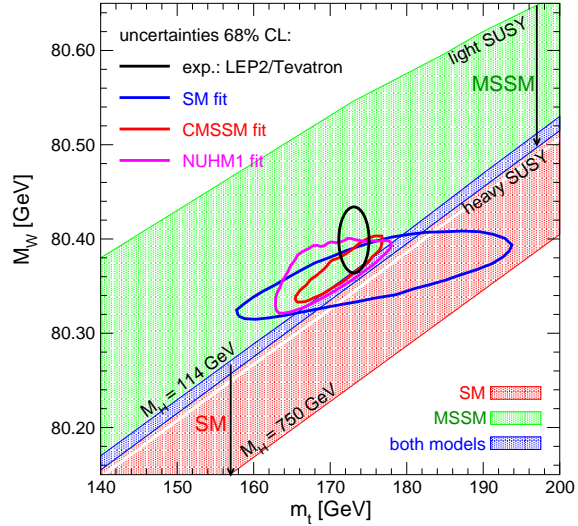}}}  
\end{center}
%%%%%%%%%%%%%%%%%%%%%%%%%%%%%%%
\vspace{-1em}
\caption{\it 
The 68\%~C.L.\ regions in the $(\mt, \MW)$ plane
predicted by a SM fit excluding the LEP Higgs constraint~\cite{lepewwg}, 
and by CMSSM and NUHM1 fits including the LEP Higgs mass constraint, 
compared with the experimental measurements from LEP2 and the Tevatron
shown as the black ellipse. 
The medium gray (red) and the dark (blue) shaded regions show the SM
prediction, foliated by lines of constant $\MHSM$ values. The light
gray (green) and the dark (blue) regions show the prediction of the
unconstrained MSSM~\cite{Heinemeyer:2006px} ranging %(top to bottom)
from light to heavy SUSY particles.}
\label{fig:MWMT09}
\end{figure}

In the SM, the precision of the confrontation between
theory and experiment is often expressed in the $(\mt, \MW)$
plane. The experimental values of these quantities are essentially
uncorrelated~\cite{lepewwg,mt1731,MW80399pm023},
\begin{align}
\mt^{\rm exp}  & = 173.1 \pm 1.3 \; \gev , \label{mtvalue} \\
\MW^{\rm exp} & = 80.399 \pm 0.023 \; \gev ,
\label{SMmtmW}
\end{align}
shown in Fig.~\ref{fig:MWMT09} as the black ellipse.
In the SM, $\mt$ is an independent input parameter, whereas the
relation between the gauge boson masses $\MW$ and $\MZ$ can be 
predicted with high precision in terms of $\mt$, the Higgs mass,
$\MHSM$, and other model parameters,
see \cite{mw2loop} and references therein. The correlation between $\mt$
and the prediction for $\MW$ is displayed in Fig.~\ref{fig:MWMT09}
(foliated by lines of
constant Higgs mass, $\MHSM$).

A fit of the SM parameters to precision observables, e.g., those
measured at the $Z$ peak~\cite{zpole}, yields indirect
predictions for $\mt$ and $\MHSM$, and hence also a
prediction for $\MW$.
The SM prediction for $\mt$ without including the experimental limits 
on $\MHSM$ and ex- or including the experimental measurement of $\MW$ 
is~\cite{lepewwg}
\begin{align}
\label{SMmtnoMW}
\mt^{{\rm fit,SM,excl.}\,\MW} &= 172.6^{+13.3}_{-10.2} \gev~, \\
\mt^{{\rm fit,SM,incl.}\,\MW} &= 179.3^{+11.6}_{-8.5} \gev~,
\label{SMmt}
\end{align}
and the SM prediction for $\MW$, excluding the experimental measurement of 
$\MW$ but either ex- or including the
experimental measurement of $\mt$ is~\cite{lepewwg}
\begin{align}
\label{SMmWnomt}
\MW^{{\rm fit,SM,excl.}\,\mt} &= 80.363 \pm 0.032 \gev~, \\
\MW^{{\rm fit,SM,incl.}\,\mt} &= 80.364 \pm 0.020 \gev~.
\label{SMmW}
\end{align}
The regions of the $(\mt, \MW)$ plane favoured at the 68\%~C.L.\ by 
direct experimental measurements (\ref{mtvalue}, \ref{SMmtmW}) 
and in the SM fit (\ref{SMmtnoMW}, \ref{SMmWnomt}), shown as the dark
(blue) contour~\cite{lepewwg-mtMW} in Fig.~\ref{fig:MWMT09}
have significant overlap, representing a non-trivial success for the
SM at the quantum level. However, we note that the overlap between the 68\%~C.L.\
contours happens in the region of Higgs mass values that are below the 
exclusion bound from the LEP SM Higgs searches, $\MHSM > 114.4
\gev$~\cite{LEPHiggsSM}, indicating a certain tension 
between the precision observables and the Higgs limit. Indeed,
the experimental central value of $\MW$ would be reached for a Higgs
mass as low as $\MHSM \sim {60} \gev$.
Combining the
indirect measurements, $\mt$ and $\MW$, 
the best-fit value of $\MHSM \sim {87} \gev$, 
and the 95\%~C.L.\ upper limit is 
$\MHSM \sim 157 \gev$~\cite{lepewwg}.
The direct searches at the Tevatron currently exclude a range
$163 \gev < \MHSM < 166 \gev$~\cite{TevHiggsSM},
as also indicated by a white line in Fig.~\ref{fig:MWMT09}, 
so that the range
$115 \gev \lsim \MHSM \lsim 150 \gev$ is favoured in a
global fit to the SM (including experimental bounds) at the 
95\%~C.L.~\cite{Gfitter}.

Turning now to our analysis in the case of supersymmetry, we note
that the prediction for $\MW$ as a function of $\mt$ in the
unconstrained MSSM gives rise to a band in 
Fig.~\ref{fig:MWMT09} (shaded green) which has only little overlap (shaded blue) 
with the band showing the range of SM predictions for Higgs masses above 
the search limit from LEP.
This is because the contribution of light supersymmetric
particles tends to increase the predicted value of $\MW$ compared to the
SM case. Furthermore, the overlap region (corresponding to the situation 
where all supersymmetric particles are heavy) is limited because, in
contrast to the SM, the value of $\Mh$ is not an independent parameter
in the MSSM, but is calculable in terms of the sparticle masses 
with an upper limit $\sim 135 \gev$~\cite{mhiggsAEC}.

We have performed fits in the CMSSM  and the NUHM1 including all
relevant experimental information as specified in~\cite{Master3}, i.e., 
we include all precision observables used in the SM fit shown in 
Fig.~\ref{fig:MWMT09} (except $\Gamma_W$, which has a minor impact)
as well as constraints from 
\gmt, flavour physics, the cold dark matter (CDM) relic density and the direct
searches for the Higgs boson and supersymmetric particles. 
The direct experimental measurements
of $\MW$ and $\mt$, on the other hand, have {\it not\/} been included in these
global fits.
The results of our fits in the CMSSM  and the NUHM1 are also displayed as
68\% C.L.\ contours in Fig.~\ref{fig:MWMT09}, and show remarkably good
agreement with the experimental measurements of $\MW$ and $\mt$.

\begin{figure}[htb!]
%%%%%%%%%%%%%%%%%%%%%%%%%%%%%%%
{\resizebox{8cm}{!}{\includegraphics{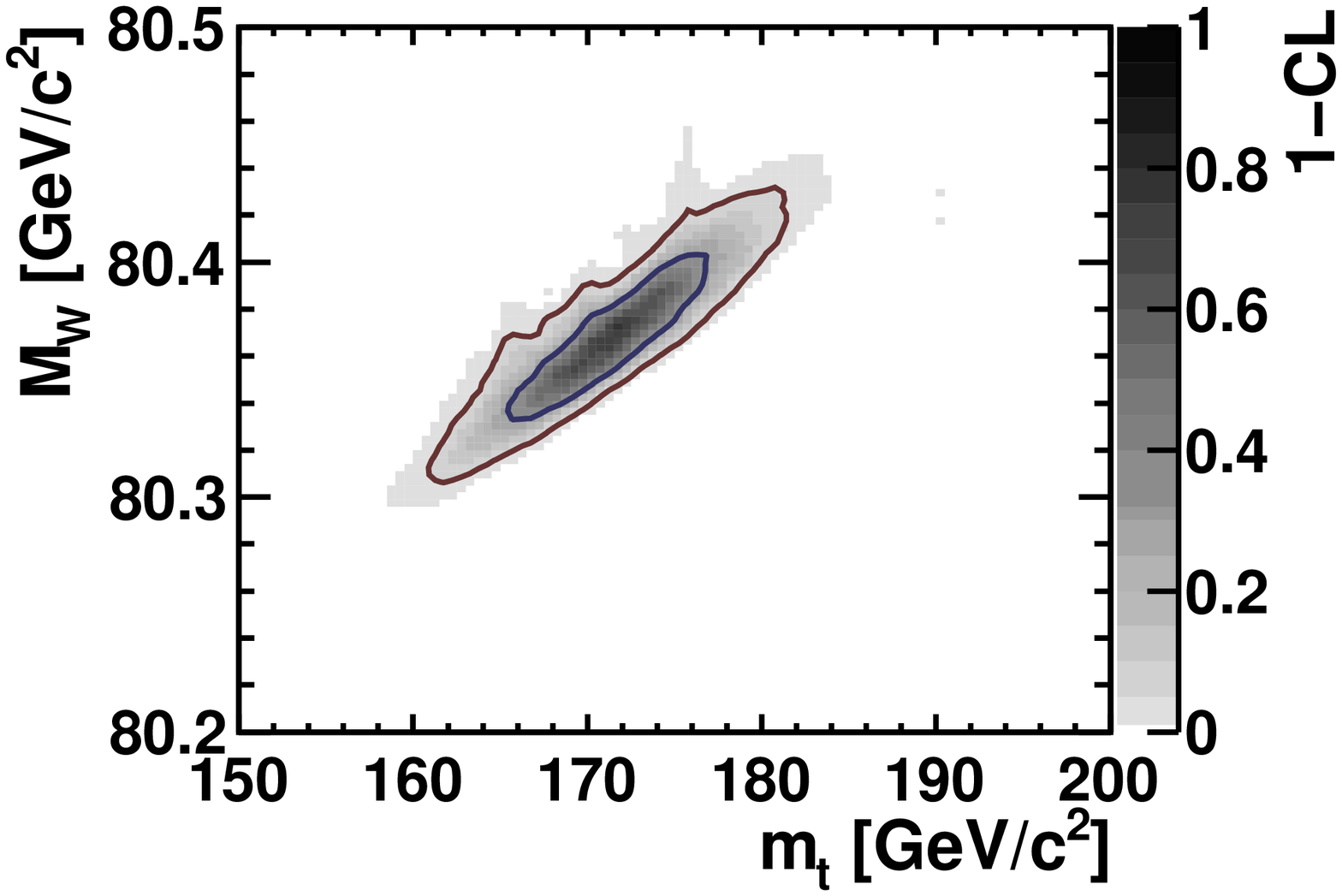}}}  
{\resizebox{8cm}{!}{\includegraphics{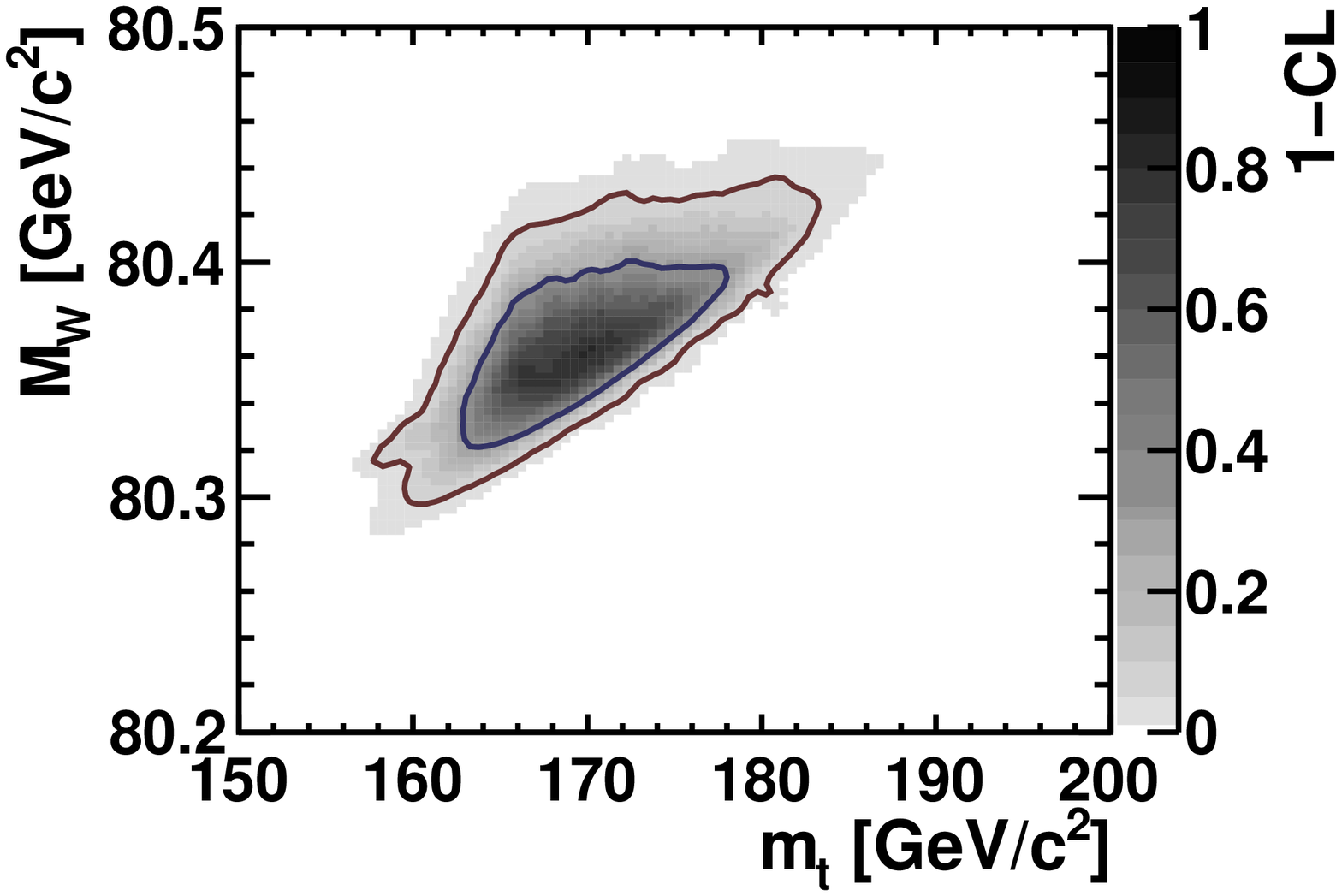}}}
%%%%%%%%%%%%%%%%%%%%%%%%%%%%%%%
\caption{\it The 68\% and 95\%~C.L.\ regions in the $(\mt, \MW)$ planes for
  the CMSSM (upper) and for the NUHM1 (lower), for fits that do not
  include the direct measurements of $\mt$ and $\MW$, but do incorporate the
  appropriate LEP constraint on $\Mh$.}
\label{fig:mtopmWLEPnomwmtop}
\end{figure}

The 68 and 95\%~C.L.\ regions in the $(\mt, \MW)$ plane
found in the CMSSM (NUHM1) fit are shown in more detail 
in the upper (lower) panel of Fig.~\ref{fig:mtopmWLEPnomwmtop}.
The fits within the MSSM differ from the SM fit in various ways. First, 
the number of free parameters is substantially larger in the MSSM, even
restricting ourselves to the CMSSM and the NUHM1. On the other hand,
more observables are included in the fits, providing extra
constraints. We recall that in the SM fits \gmt\ and the $B$-physics
observables have a minor impact on the best-fit regions, and are
not included in the results shown above, which are taken from 
\cite{lepewwg} (see e.g.\ \cite{pdg} for an alternative approach), while
the relic density of CDM cannot be accommodated in the SM.
Furthermore, as already noted, whereas the light Higgs boson
mass is a free parameter 
in the SM, it is a function of the other parameters in the
CMSSM and NUHM1. In this way, for example, the masses of the scalar tops and
bottoms enter not only directly into the prediction of the various
observables, but also indirectly via their impact on $\Mh$.
This provides additional motivation for including
the experimental constraints on $\Mh$ into the fits in the MSSM.

%%%%%%%%%%%%%%%%%%%%%% F I G U R E %%%%%%%%%%%%%%%%%%%%%%%%%%%%%%%%%%%
%%%%%%%%%%%%%%%%%%%%%% F I G U R E %%%%%%%%%%%%%%%%%%%%%%%%%%%%%%%%%%%

\begin{figure}[htb!]
%%%%%%%%%%%%%%%%%%%%%%%%%%%%%%%
{\resizebox{8cm}{!}{\includegraphics{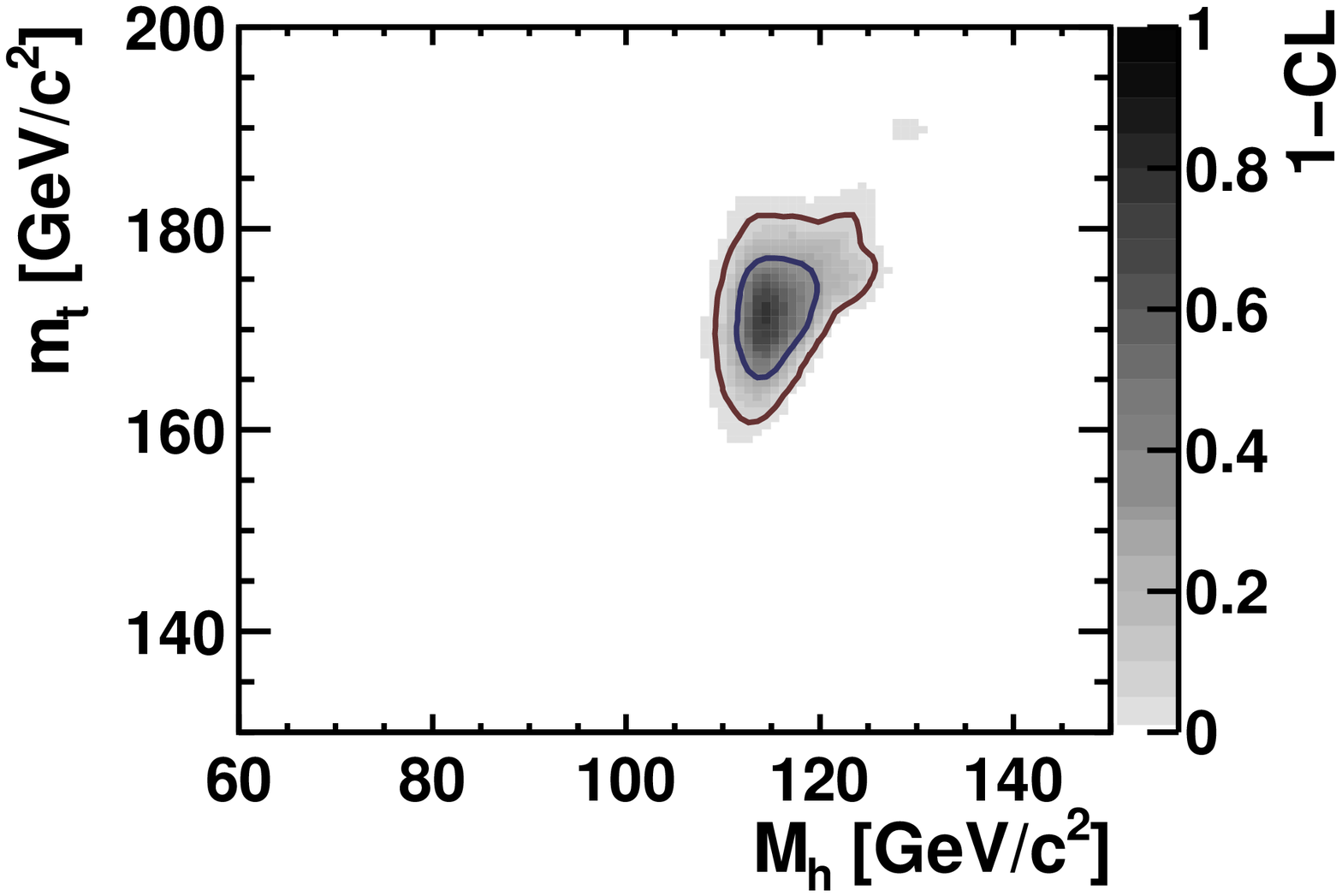}}}  
{\resizebox{8cm}{!}{\includegraphics{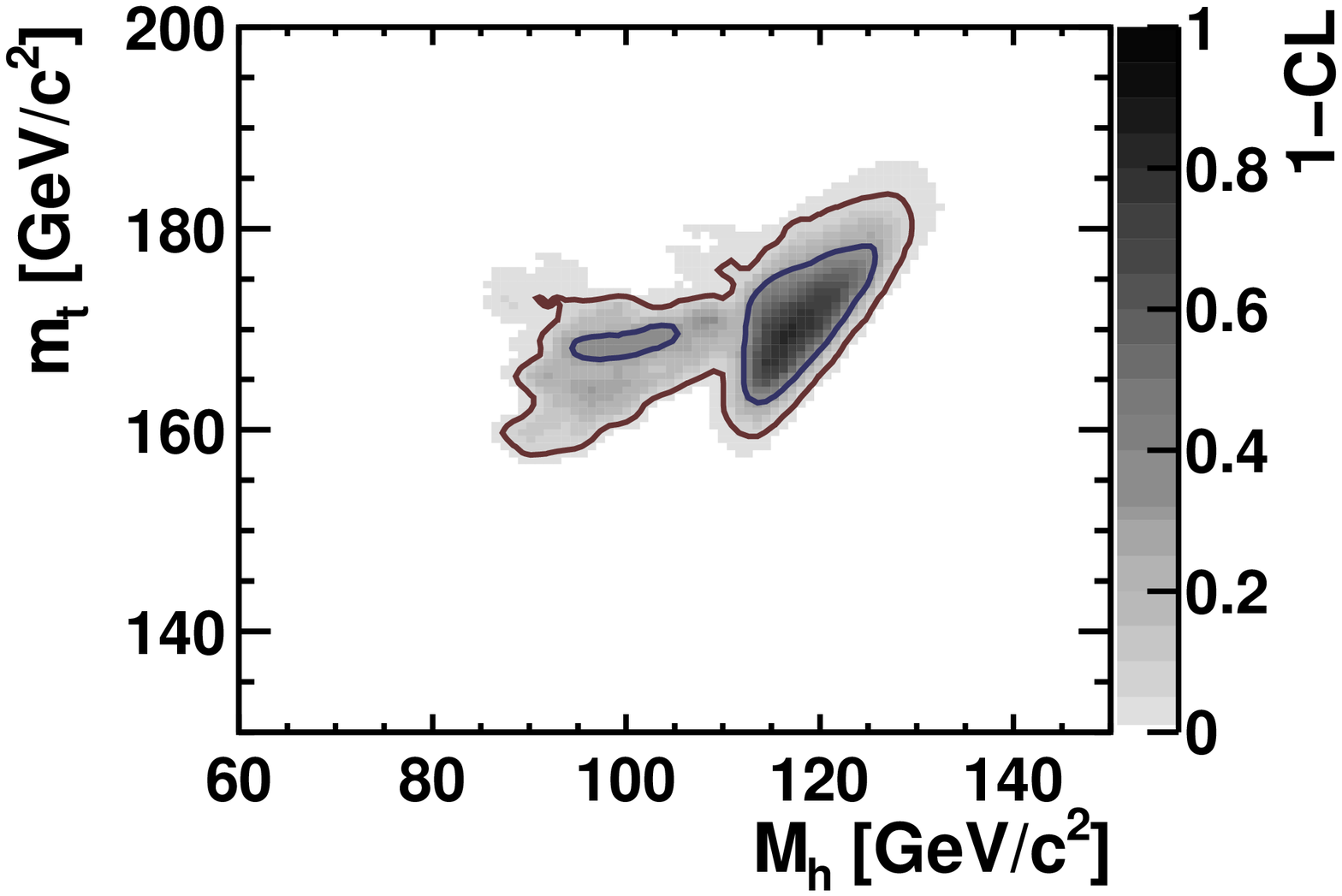}}}
%%%%%%%%%%%%%%%%%%%%%%%%%%%%%%%
\caption{\it The 68\% and 95\%~C.L.\ regions in the $(\Mh, \mt)$ planes for
  the CMSSM (upper plot) and for the NUHM1 (lower plot), for fits that do not
  include the direct measurements of $\mt$ and $\MW$, but do incorporate the
  appropriate LEP constraint on $\Mh$.}
\label{fig:mtopmhLEPnomwmtop}
\end{figure}

In Fig.~\ref{fig:mtopmhLEPnomwmtop}, we show the results of the same fit as
in Fig.~\ref{fig:mtopmWLEPnomwmtop}, but now in the $(\Mh,\mt)$~plane for the
CMSSM (NUHM1) in the upper (lower) panel.
The LEP lower limit of 114~GeV is applicable
in the CMSSM~\cite{Ellis:2001qv,Ambrosanio:2001xb}, but cannot
always be directly applied in the NUHM1, since 
there are regions of the NUHM1 parameter space where the $hZZ$ coupling 
is suppressed relative to its value in the 
SM~\cite{LEPHiggsMSSM}. We
use the prescription given in~\cite{Master3} to calculate the $\chi^2(\Mh)$ 
contribution for points with suppressed $hZZ$ couplings, and see in
the lower panel of Fig.~\ref{fig:mtopmhLEPnomwmtop} a significant set of
NUHM1 points with $\Mh \ll 114 \gev$: these reflect the shape of the $\Delta \chi^2$
function in the right panel of Fig.~4 of~\cite{Master3}. The appearance of a local minimum at $\Mh \approx 100 \gev$ in the lower plot of Fig.~\ref{fig:mtopmhLEPnomwmtop}
is statistically not significant. It sensitively depends on
the details of the implementation of the Higgs search bounds.

%%%%%%%%%%%%%%%%%%%%%% F I G U R E %%%%%%%%%%%%%%%%%%%%%%%%%%%%%%%%%%%
%%%%%%%%%%%%%%%%%%%%%% F I G U R E %%%%%%%%%%%%%%%%%%%%%%%%%%%%%%%%%%%

\begin{figure}[htb!]
%%%%%%%%%%%%%%%%%%%%%%%%%%%%%%%
{\resizebox{8cm}{!}{\includegraphics{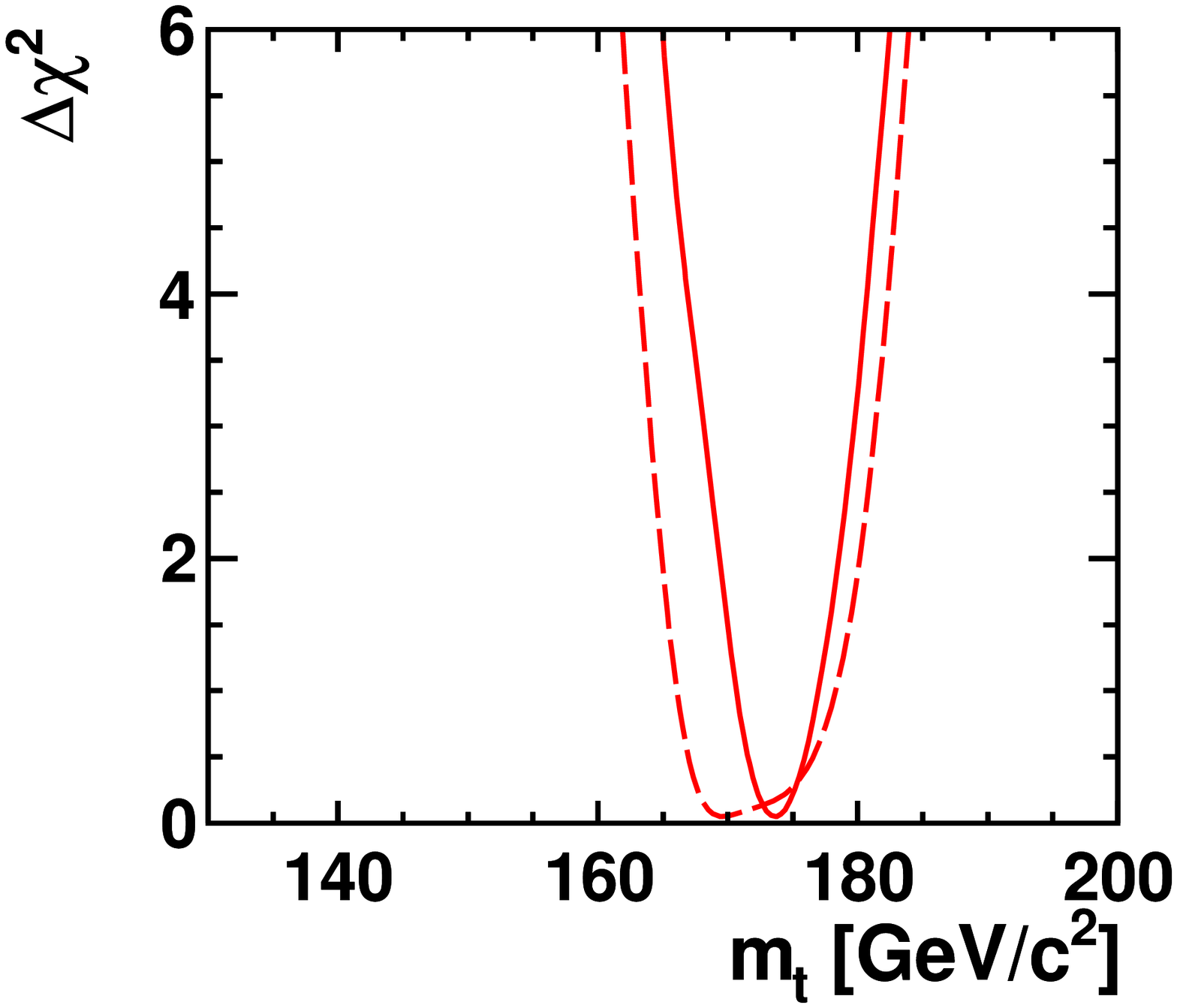}}}  
{\resizebox{8cm}{!}{\includegraphics{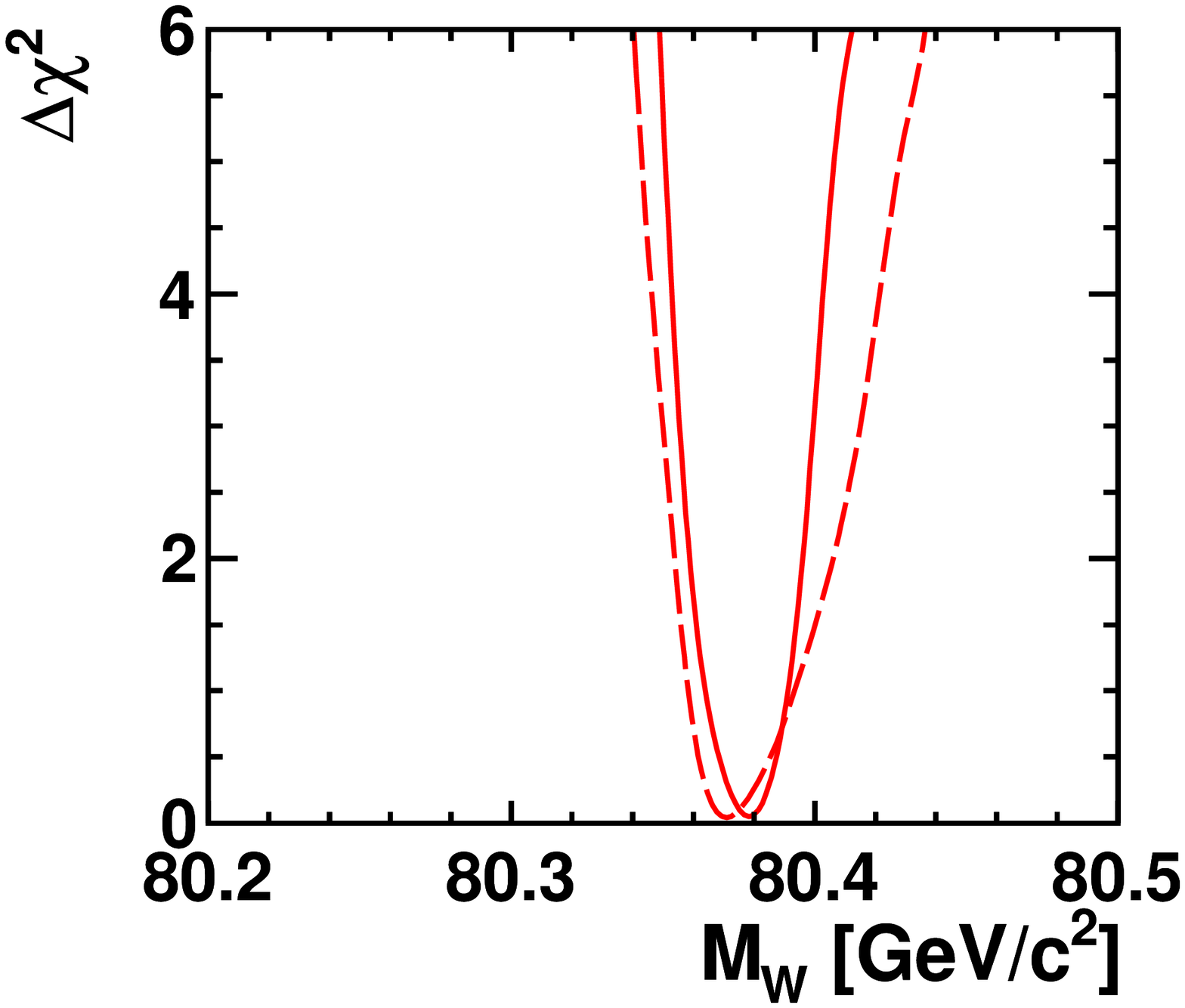}}}
%%%%%%%%%%%%%%%%%%%%%%%%%%%%%%%
\caption{\it The $\chi^2$ functions for $\mt$ (upper panel) in the CMSSM (solid)
  and NUHM1 (dashed) excluding the direct $\mt$ mass measurement but
  including all the other experimental information. The corresponding  $\chi^2$ functions for $\MW$
(lower panel) excluding the direct $\MW$ mass measurement but again
  including all the other experimental information.}  
\label{fig:LLmt}
\end{figure}

We now turn to the single-variable $\chi^2$ functions for $\mt$
and $\MW$.
In the upper panel of  Fig.~\ref{fig:LLmt}, we show the $\chi^2$
functions for $\mt$ in the CMSSM 
and NUHM1 as solid and dashed lines respectively with $\MW$ included in
the fit (as before, the direct measurement of $\mt$ is not included
in this fit). Comparing the results with 
the SM fit, we find that these rise more
sharply, in particular for larger values of $\mt$, than they 
would in the SM fit, indicating that the upper bound on $\mt$ from the
indirect prediction in the MSSM is significantly reduced compared to the SM case.
We find the 68\%~C.L.\ ranges
\begin{align}
\mt^{{\rm fit,CMSSM,incl.}\,\MW}  &=  173.8^{+3.2}_{-3.1} \gev , \\
\mt^{{\rm fit,NUHM1,incl.}\,\MW} &= 169.5^{+8.8}_{-3.4} \gev .
\label{mtnomtmWLEP}
\end{align}
Comparing with the SM fit result (\ref{SMmt}), we find lower central
values for $\mt$ in both the CMSSM and NUHM1 in better agreement with
 the experimental result (\ref{mtvalue}). 
The reduction in the upper bound on $\mt$ reflects, in particular,
the fact that the additional contribution from the ${\tilde t}$ and ${\tilde b}$
enters with the same sign as the leading
SM-type contribution to the precision observables that is proportional
to $\mt^2$. A non-vanishing contribution from superpartners therefore
tends to reduce the preferred value of $\mt$ compared to the SM fit. 
It should be noted in this context that the smaller uncertainties in 
$\mt$ found in the supersymmetric fits compared to the SM case 
(particularly in the CMSSM) can in part also be 
attributed to the fact that a 
larger set of observables has been used in the CMSSM and NUHM1 fits.

%%%%%%%%%%%%%%%%%%%%%% F I G U R E %%%%%%%%%%%%%%%%%%%%%%%%%%%%%%%%%%%
%%%%%%%%%%%%%%%%%%%%%% F I G U R E %%%%%%%%%%%%%%%%%%%%%%%%%%%%%%%%%%%

For the $W$~boson mass, we find the $\chi^2$ functions including $\mt$
in the fit in the CMSSM (solid) and NUHM1 (dahed)
shown in the lower panel of Fig.~\ref{fig:LLmt}, and the
corresponding 68\%~C.L.\ ranges
\begin{align}
\MW^{{\rm fit,CMSSM,incl.}\,\mt}  &= 80.379^{+0.013}_{-0.014} \gev ,  \\
\MW^{{\rm fit,NUHM1,incl.}\,\mt} &= 80.370^{+0.024}_{-0.011} \gev .
\label{mWnomWmtLEP}
\end{align}
The best-fit values of these predictions are substantially higher
than the SM prediction (\ref{SMmW}) based on precision electroweak data
(in particular in the CMSSM)
and are closer to the 
experimental value (\ref{SMmtmW}), again with smaller uncertainties.

\begin{figure}[htb!]
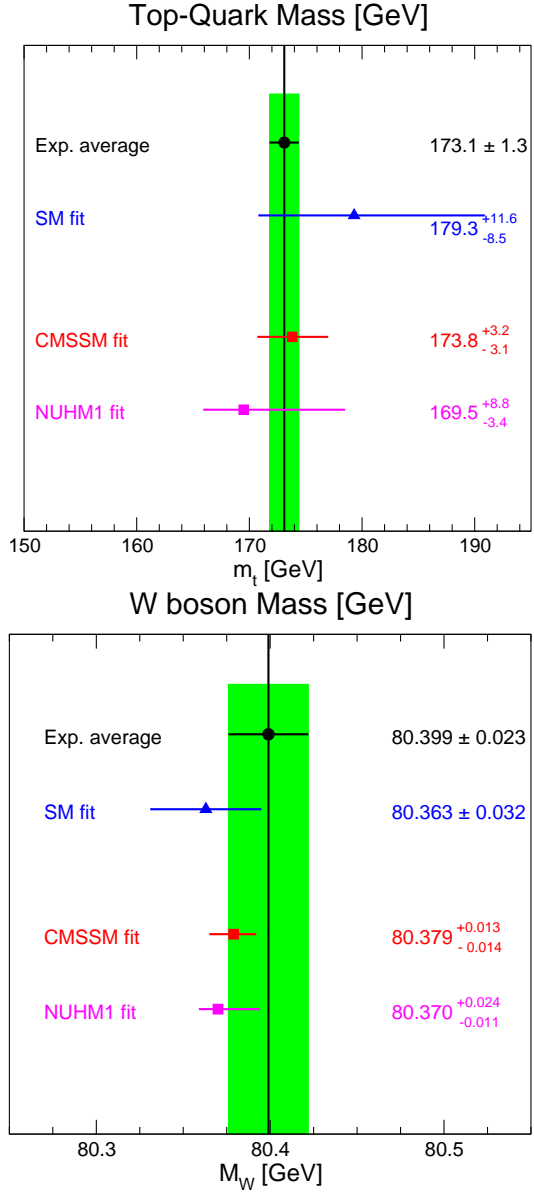

%%%%%%%%%%%%%%%%%%%%%%%%%%%%%%%
{\resizebox{7cm}{!}{\includegraphics{MT09_MC_02.eps}}}  
{\resizebox{7cm}{!}{\includegraphics{MW09_MC_02.eps}}}
%%%%%%%%%%%%%%%%%%%%%%%%%%%%%%%
\caption{\it The 68\%~C.L.\ ranges for $\mt$ (upper panel) and $\MW$ (lower
  panel) including (from top to bottom) the experimental average, and the
  predictions of the SM (not incl.\ the $\MHSM$
    limits)~\cite{lepewwg}, CMSSM and NUHM1 
  fits, using all the available information except the direct mass
  measurement.} 
  %\htb{and excluding or including the $\Mh$ bounds}.} 
\label{fig:final}
\end{figure}

We summarize our main results in Fig.~\ref{fig:final}.
The upper (lower) panel compares the experimental measurement of
$\mt$ ($\MW$) with the predictions of a SM fit to precision electroweak
data and our final predictions in the CMSSM and NUHM1.
{\it The resulting agreement of the final 
predictions for $\mt$ with the experimental value (\ref{mtvalue}) is
remarkable, almost embarrassingly good in the CMSSM case}, and very
good in the NUHM1.
Compared to the SM fit, the best-fit values for $\MW$ in the CMSSM and
NUHM1 are closer to the  
experimental value (\ref{SMmtmW}), and in the CMSSM case the best-fit
value lies within the experimental 68\%~C.L.\ range. We conclude that the CMSSM
and NUHM1 pass with flying colours the test of reproducing the successful
SM predictions of $\mt$ and $\MW$, even improving on them. We can only hope that
this probe of SUSY at the loop level will soon be made even more precise with
the discovery of sparticles at the LHC.

This work was supported in part by the European Community's Marie-Curie
Research Training Network under contracts MRTN-CT-2006-035505
`Tools and Precision Calculations for Physics Discoveries at Colliders'
and MRTN-CT-2006-035482 `FLAVIAnet', and by the Spanish MEC and FEDER under 
grant FPA2005-01678. The work of S.H. was supported 
in part by CICYT (grant FPA~2007--66387), and
the work of K.A.O. was supported in part
by DOE grant DE--FG02--94ER--40823 at the University of Minnesota.

%%%%%%%%%%%%%%%%%%%%%% F I G U R E %%%%%%%%%%%%%%%%%%%%%%%%%%%%%%%%%%%
%%%%%%%%%%%%%%%%%%%%%% F I G U R E %%%%%%%%%%%%%%%%%%%%%%%%%%%%%%%%%%%

%---------------------------------------------------------------------

\end{document}